\documentclass[letterpaper]{article} 
\usepackage{aaai2026}  
\usepackage{times}  
\usepackage{helvet}  
\usepackage{courier}  
\usepackage[hyphens]{url}  
\usepackage{graphicx} 
\usepackage{natbib}  
\usepackage{caption} 
\usepackage{algorithm}
\usepackage{algorithmic}
\usepackage{graphicx}
\usepackage{multirow}
\usepackage{booktabs}
\usepackage{amsfonts}   

\usepackage{amsmath}
\usepackage[table]{xcolor}
\definecolor{coral}{HTML}{F08080}

\definecolor{blueaccent}{HTML}{0070BB}
\definecolor{coral}{HTML}{F08080}
\usepackage{newfloat}
\usepackage{listings}
\DeclareCaptionStyle{ruled}{labelfont=normalfont,labelsep=colon,strut=off} 
\floatstyle{ruled}
\newfloat{listing}{tb}{lst}{}
\floatname{listing}{Listing}
\title{Towards Neural Audio Codec Source Parsing}
\author{
    Orchid Chetia Phukan\textsuperscript{\rm 1}\equalcontrib,
    Girish\textsuperscript{\rm 1,2} \equalcontrib,
    Mohd Mujtaba Akhtar\textsuperscript{\rm 1,3} \equalcontrib,\\
    Arun Balaji Buduru\textsuperscript{\rm 1},
    Rajesh Sharma\textsuperscript{\rm 4,5}
}
\affiliations{
    IIIT-Delhi, India\textsuperscript{\rm 1},
    UPES, India\textsuperscript{\rm 2}, 
    V.B.S.P.U, India\textsuperscript{\rm 3},
    University of Tartu, Estonia\textsuperscript{\rm 4},
    Plaksha University, India\textsuperscript{\rm 5}

    \textbf{Correspondence:} {orchidp@iiitd.ac.in}
}

\begin{document}

\maketitle

\begin{abstract}
A new class of audio deepfakes—codecfakes (CFs)—has recently caught attention, synthesized by Audio Language Models that leverage neural audio codecs (NACs) in the backend. In response, the community has introduced dedicated benchmarks and tailored detection strategies. As the field advances, efforts have moved beyond binary detection toward source attribution, including open-set attribution, which aims to identify the NAC responsible for generation and flag novel, unseen ones during inference. This shift toward source attribution improves forensic interpretability and accountability. However, open-set attribution remains fundamentally limited: while it can detect that a NAC is unfamiliar, it cannot characterize or identify individual unseen codecs. It treats such inputs as generic ``unknowns'', lacking insight into their internal configuration. This leads to major shortcomings: limited generalization to new NACs and inability to resolve fine-grained variations within NAC families. To address these gaps, we propose Neural Audio Codec Source Parsing (NACSP) - a paradigm shift that reframes source attribution for CFs as structured regression over generative NAC parameters such as quantizers, bandwidth, and sampling rate. We formulate NACSP as a multi-task regression task for predicting these NAC parameters and establish the first comprehensive benchmark using various state-of-the-art speech pre-trained models (PTMs). To this end, we propose \texttt{\textbf{HYDRA}}, a novel framework that leverages hyperbolic geometry to disentangle complex latent properties from PTM representations. By employing task-specific attention over multiple curvature-aware hyperbolic subspaces, \texttt{\textbf{HYDRA}} enables superior multi-task generalization. Our extensive experiments show \texttt{\textbf{HYDRA}} achieves top results on benchmark CFs datasets compared to baselines operating in Euclidean space.

\end{abstract}


\section{Introduction}

\label{intro}
\textit{``If seeing is no longer believing, then hearing certainly isn’t either.''} Once a concern for visual domain, this message now holds ground in the audio domain too. Driven by progress in generative modeling of voices, audio deepfakes (ADs) have grown from mere technical methodologies into potential tools with significant real-world implications. These synthetic voices synthesized by neural architectures convincingly mimic human speech—capturing not only linguistic content but also prosody, accent, and emotional tone. While they offer benefits in accessibility and creative media, their malicious use—such as impersonating individuals in scams and spreading misinformation—poses serious risks. The most common types of ADs are produced using Text-to-Speech (TTS) and Voice Conversion (VC) systems \cite{nautsch2021asvspoof, liu2023asvspoof}. \par

Another growing area involves singing voice ADs generated through singing voice synthesis and singing voice conversion models \cite{zang2024singfake, zang24_interspeech}. These are widely used in creating cover songs and virtual performances, but also raise questions about originality and consent. In response, detection of such ADs demands robust, promising solutions and an need of the hour. Early approaches to detecting ADs primarily focused on binary classification—distinguishing between real and fakes—using handcrafted features \cite{patel15_interspeech, todisco2017constant, 7472724} or even raw audio signals. Following this, the landscape of audio deepfake detection (ADD) has changed from handcrafted features to the use of pre-trained models (PTMs). These PTMs are preferred as they provide performance benefit as well as preventing training models from scratch. As such previous works have explored different state-of-the-art (SOTA) PTMs like WavLM, Unispeech-SAT, Wav2vec2, Whisper, XLS-R for ADD \cite{kawa23b_interspeech, 10.1145/3664647.3681345, chetia-phukan-etal-2024-heterogeneity, li2024cross}.  \par
\begin{figure}[!h]
    \centering
    \includegraphics[width=0.6\linewidth]{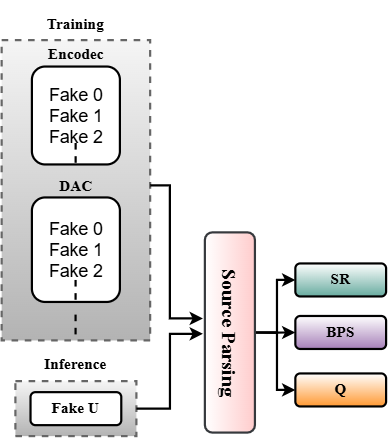}
    \caption{Demonstration of NACSP; Fake U represents a unknown audio generated by an unseen NAC; SR: Sampling Rate, BPS: Bits per second, Q: Quantizers}
    \label{nacsp}
\end{figure}
Recently, a new and more subtle category of ADs has emerged—codecfake (CF) \cite{wu24p_interspeech, lu24f_interspeech}—generated by Audio Language Models (ALMs) which utilizes neural audio codecs (NACs) in backend for tokenizing and reconstructing audio. In contrast to conventional speech synthesis pipelines that use vocoders for waveform production, CF are generated by a discrete compression-decompression process in which speech is compressed into latent tokens through a NAC (e.g., EnCodec \cite{defossezhigh}, SoundStream \cite{zeghidour2021soundstream}), represented by an ALM, and decoded back into waveforms. This codec-based method allows for highly realistic audio production. Due to the difference in the generation process of CF, ADD models trained on vocoder synthesized data shows lower detection performance when exposed to CF. As a remedy, researchers have released benchmark CF datasets for the community to build on \cite{wu24p_interspeech, lu24f_interspeech} and have developed various approaches for detection of CF. These works have shown the ADD models trained on CF datasets, shows improved performance in detecting CF. Further,  \citealt{10830534} also explored various PTMs such as WavLM, Wav2vec2 with LCNN, AASIST as downstream network for detecting CF. In this work, we specifically focus on CF. \par

As ADD research space continues to advance, research is shifting beyond traditional binary classification toward source attribution or source tracing—identifying not just whether audio is fake, but pinpointing the specific generative model responsible whether it be NAC, TTS or VC system. This transition represents a pivotal step in enhancing the interpretability, transparency, and forensic utility of ADD systems, making them more actionable for real-world and policy-driven scenarios. In response, researchers have proposed a variety of approaches to enable reliable source attribution \cite{yan2022system, 10889868}. To further strengthen source attribution systems under open-set conditions—where the generative source encountered at inference was not seen during training—several recent works have begun to explore this challenging direction \cite{bhagtani2024attribution, xie24_interspeech, 10889399}. Complementing this direction, \citealt{xie2025neural} advanced source attribution for CFs, addressing a pressing need in the evolving landscape of audio deepfake forensics. They handle both in-distribution and open-set scenarios.\par

However, open-set attribution remains fundamentally limited: while it can flag a NAC as unfamiliar, it fails to characterize or pinpoint which unseen NAC generated the audio treating all novel sources as generic ``unknown''. It also doesn't provide insight into their internal generative configuration. This leads to three major shortcomings: limited generalization to novel NACs, inability to resolve fine-grained variations within codec families, and poor forensic interpretability. To address these gaps, we propose NAC source parsing (NACSP) (Figure \ref{nacsp})—a paradigm shift that reframes source attribution for CF as structured regression over NAC parameters (e.g., quantizers, bits per second, sampling rate) in a multi-task setup rather than discrete classification. Through estimation of these NAC parameters, we will be able to estimate the NAC through which the audio was generated form. Notably, NACSP unifies in-distribution and open-set attribution into a single regression modeling formulation, eliminating the need for separate modeling or post hoc open-set detection modules. \par

To establish NACSP, we first present a comprehensive evalaution of diverse set of SOTA PTMs. As benchmark various SOTA PTMs for NACSP, as they have shown potential for related tasks such as source attribution \cite{klein24_interspeech} and ADD \cite{chetia-phukan-etal-2024-heterogeneity}. However, given that PTMs representations entangles different acoustic cues within their representational space and each NAC parameter may rely on different subsets of acoustic features—modeling all parameters jointly in a shared space can result in representational interference and suboptimal performance. To address this, we propose \texttt{\textbf{HYDRA}} (\texttt{\textbf{HY}}perbolic Attention for \texttt{\textbf{D}}isentangled \texttt{\textbf{R}}epresent\texttt{\textbf{A}}tion Modeling) —a novel framework that employs task-specific attention over multiple hyperbolic subspaces, each optimized to capture the latent structure relevant to a specific NAC parameter. By separating the learning dynamics for each NAC attribute, \texttt{\textbf{HYDRA}} effectively disentangles the features from PTMs representational space and enables more precise estimation. Across all benchmarks, \texttt{\textbf{HYDRA}} consistently outperforms competitive baseline models that dwells in euclidean space, establishing SOTA for NACSP. Our work opens a new research direction in ADD forensics and encourages the community to explore structured, interpretable modeling of generative audio traces.\par

\noindent \textbf{To summarize, the main contributions of the paper are as follows:}\par
\noindent (i) We introduce NACSP, a novel paradigm that reframes NAC source attribution as structured regression over NAC parameters rather than discrete classification. NACSP unifies in-distribution and open-set attribution within a single modeling framework, addressing key limitations in existing forensic methods. (ii) We present the first extensive benchmark evaluating diverse SOTA PTMs on the NACSP task. We perform our experiments on benchmark CF datasets such as ST-codecfake \cite{xie2025neural} and CodecFake \cite{wu24p_interspeech}. (iii) We propose \texttt{\textbf{HYDRA}}, a novel framework that employs task-specific attention over multiple hyperbolic subspaces, each optimized for a particular NAC parameter. HYDRA effectively disentangles relevant acoustic factors responsible for each NAC parameter from the PTMs representational space, significantly outperforming baseline models, and establishing SOTA results in NACSP. \textit{We will release the source code and model checkpoints upon completion of double-blind review to facilitate future research for NACSP.}

\section{Related Works}

\noindent In this section, we review key works in source attribution that lays the groundwork for NACSP. Early studies introduced the task of identifying attacker signatures, demonstrating that RNN-based representations could reliably distinguish both seen and unseen attackers \cite{muller22b_interspeech}. Building on this, \citealt{yan2022initial} highlighted the potential of vocoder-specific fingerprints as unique identifiers for source characterization. More recently, \citealt{klein24_interspeech} explored the use of SOTA PTMs such as Wav2vec2 and Whisper, showcasing their effectiveness in improving source attribution accuracy. These works however, haven't considered audios synthesized from diffusion-based synthesis methods, so \citealt{bhagtani2024attribution} took the first step for source attribution of diffusion-based sources. They have experimented with various baseline methods such as MFCC-ResNet, Spec-ResNET, Wav2vec2 and so on both for in-distribution sources as well as open-set scenarios. \citealt{mishra2025towards} performs ADD and closed-set attribution where the source lies within the training set using probabilistic attribute embeddings derived from high-level AD generation metadata (e.g., vocoder type, input format). An benefit of this framework is that it introduces explainability via shapley value analysis for each attribute involved in the generation process.
\citealt{xie24_interspeech} proposed a two-stage framework for ADD and attribution that also supports open-set attribution by considering it as a novel class. The framework first learns a real-vs-fake discriminator using OC-Softmax, then applies a classifier in addition with open-set source detector to trace the source to a known algorithm or detect it as novel. However, all these studies were on vocoder-based synthesis methods and none focusing on CF source attribution and to solve this gap, \citealt{xie2025neural} took the first step. For identifying a specific NAC used and even when the NAC was unseen during training, they achieve this by learning discriminative features from multiple NAC types and applying threshold-based open-set detection. However, they treat each NAC as a discrete class and all open-set NACs are grouped together to a single unknown class. This approach lacks the ability to pinpoint or interpret the exact configuration of an unseen NAC. To address this, we propose NACSP task in our work—a shift from classifying NACs to regressing key NAC parameters. This enables fine-grained attribution even for unseen NACs.

\section{Pre-trained Models}
In this section, we discuss the PTMs under consideration in our work. We select these PTMs as they are SOTA in their respective benchmarks and have shown SOTA performance in various speech processing tasks. Unispeech-SAT\footnote{\url{https://huggingface.co/microsoft/unispeech-sat-base-100h-libri-ft}} \cite{chen2022unispeech} improves speech representation learning by incorporating speaker-awareness into self-supervised training through a multi-task framework. It is a SOTA PTM in SUPERB benchmark. WavLM \cite{chen2022wavlm}\footnote{\url{https://huggingface.co/microsoft/wavlm-base}} is a self-supervised learning speech PTM trained for speech masked modeling and denoising simultaneously. It is also a SOTA PTM in SUPERB. Wav2vec2\footnote{\url{https://huggingface.co/facebook/wav2vec2-base}} \cite{baevski2020wav2vec} is trained in a self-supervised manner to solve a contrastive task by leveraging quantized representations of jointly learned latent features. We make usage of WavLM, Unispeech-SAT, Wav2vec2 base version with 94.70M, 94.68M, 95.04M parameters respectively and pre-trained on librispeech 960 hours english data. We also consider SOTA multilingual PTMs such as XLS-R \cite{babu22_interspeech}, Whisper \cite{radford2023robust} and MMS \cite{pratap2024scaling} which are SOTA in different multilingual speech processing tasks in different benchmarks. XLS-R\footnote{\url{https://huggingface.co/facebook/wav2vec2-xls-r-1b}}, Whisper\footnote{\url{https://huggingface.co/openai/whisper-base}} , and MMS\footnote{\url{https://huggingface.co/facebook/mms-1b}} are pre-trained on 128, 96, and 1400 languages, respectively. While XLS-R and MMS are based on the Wav2vec2 architecture, Whisper adopts a standard transformer-decoder design. In our experiments, we utilize the 300M, 74M, and 1B parameter versions of XLS-R, Whisper, and MMS, respectively. Additionally, we also consider speaker recognition PTMs such as x-vector\footnote{\url{https://huggingface.co/speechbrain/spkrec-xvect-voxceleb}} \cite{8461375} ECAPA\footnote{\url{https://huggingface.co/speechbrain/spkrec-ecapa-voxceleb}} \cite{desplanques20_interspeech} which has shown effectiveness for ADD \cite{chetia-phukan-etal-2024-heterogeneity}. Both x-vector and ECAPA are time-delay neural networks trained in a supervised manner on a combined dataset of VoxCeleb1 and VoxCeleb2. ECAPA offers improved speaker recognition performance over x-vector, which itself outperforms the earlier i-vector approach. The x-vector model comprises approximately 4.2M parameters. \par

All the audio samples are resampled to 16kHz before passing as input to the PTMs. We extract representations by applying average pooling over the last hidden state of each frozen PTM. For Whisper, the encoder outputs are used as representations by discarding the decoder. The resulting representation dimensions extracted from the PTMs are as follows: 768 for UniSpeech-SAT, WavLM, Wav2vec2; 512 for x-vector and Whisper; 192 for ECAPA; and 1280 for MMS and XLS-R.

\begin{figure}[hbt!]
    \centering
    \includegraphics[width=0.7\linewidth]{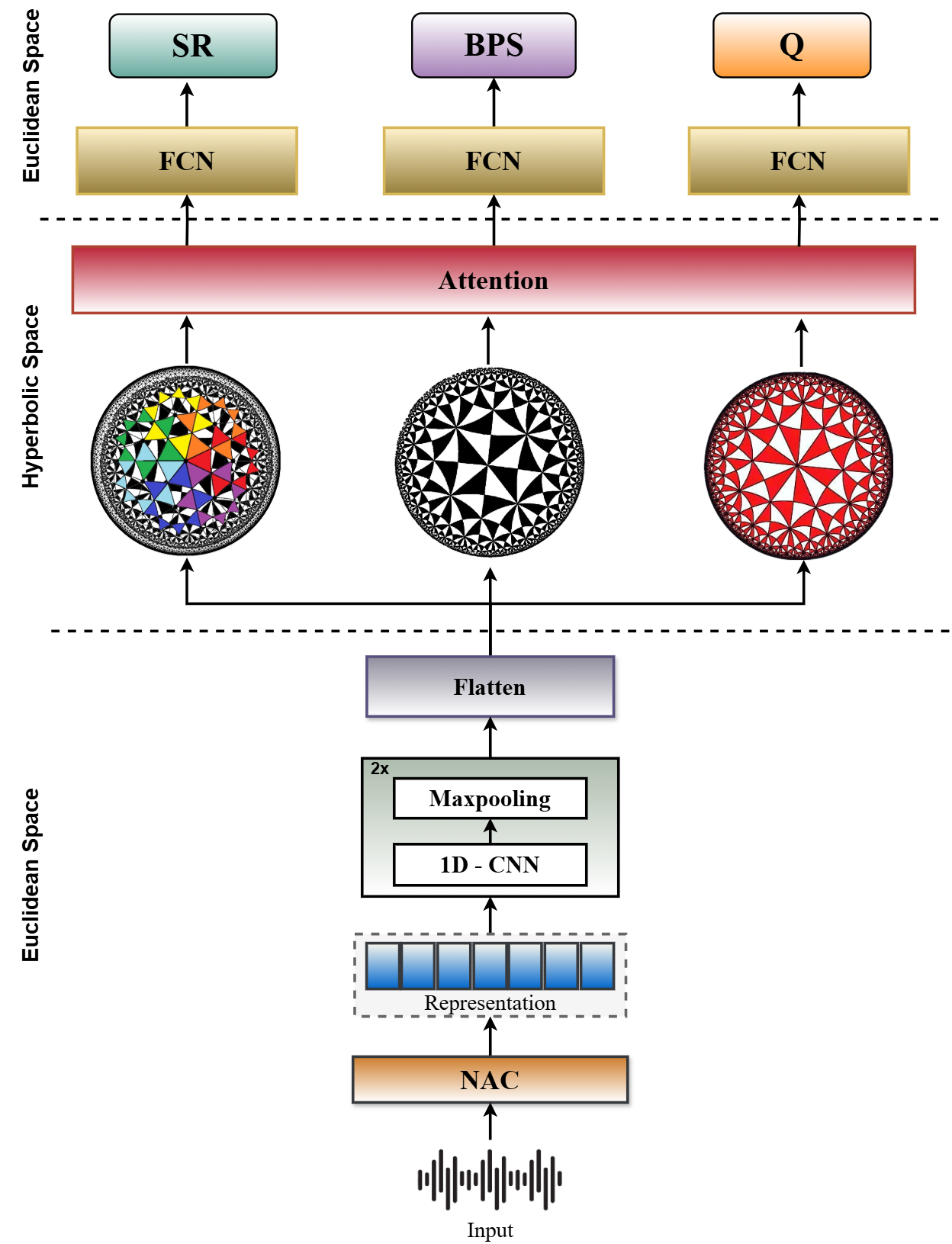}
     \caption{\textbf{\texttt{HYDRA}}; FCN stands for fully connected network}
    \label{proposedhyperbolic}
\end{figure}

\section{Modeling}

\subsection{Baseline Modeling}
\label{baseline}
In this section, we describe the baseline modeling framework using various PTMs. We adopt a CNN-based architecture as the downstream network, motivated by its widespread use and effectiveness across multiple speech processing tasks \cite{chetiaphukan23_interspeech, rathod23_interspeech, chetia-phukan-etal-2024-heterogeneity}. The extracted representations are first passed through a convolutional block comprising two 1D CNN layers with 64 and 128 filters, respectively, both using a kernel size of 3. The output is then flattened and routed through three parallel branches, each dedicated to predicting one of the three NAC parameters. Each task-specific branch consists of a fully connected network with two dense layers of 120 and 30 neurons using ReLU activation, followed by an output layer with linear activation to perform regression.

\subsection{HYDRA}
In this section, we introduce our proposed framework, \texttt{HYDRA}, for NACSP, as illustrated in Figure~\ref{proposedhyperbolic}. PTMs representational space encode rich information that capture various aspects of acoustic cues such as intonation, tone, rhythm, and speaker identity. However, these information in representations are highly entangled in a way that makes it difficult to extract what’s relevant for different task in a multi-task format. NACSP involves multiple tasks—such as predicting the sampling rate, bits per second, and quantizers—each of which depends on different parts of this embedded information. For example, sampling rate prediction relies on how finely time is captured in the audio, bits per second prediction depends on how much detail or loss the compression has introduced, and quantizers prediction relates to identifying the number of distinct building blocks used during encoding. Although these hints lie in PTM representations, they are not individually available as a result of entanglement. Moreover, the information encoded in PTM representations is inherently hierarchical from low-level acoustics to high-level semantics. This hierarchical structure motivates us to for the usage of hyperbolic geometry. Hyperbolic spaces, due to their negative curvature, are especially well-suited to represent structured or hierarchical data—and offer greater representational capacity compared to flat (Euclidean) spaces. As such, we propose \textbf{\texttt{HYDRA}}, which maps the joint latent representational space of PTMs onto three individual hyperbolic subspaces corresponding to each NACSP task. Essentially, projecting into distinct subspaces enables each NACSP task to learn from its own dedicated space without interference from others and will result in more effective predictions. The detailed flow of \textbf{\texttt{HYDRA}} is as follows: The input PTM representations are passed through a shared convolutional block with the same architectural details as in Section~\ref{baseline}. The flattened vector \( z \) is then projected into multiple hyperbolic subspaces, each assigned to a NACSP task \( k = 1, \ldots, K \). Each subspace is modeled as a poincaré ball \(\mathbb{H}^d\) with learned curvature \( c_k > 0 \). The projection is performed via a trainable linear map followed by the poincaré exponential map at the origin: \( z^{(k)} = \exp_0^{c_k}(z)\), where \( \exp_0^{c_k} : \mathbb{R}^d \to \mathbb{H}^d \) is the poincaré exponential map that maps euclidean vectors into the hyperbolic poincaré ball of curvature \( c_k \). The exponential map \(\exp_0^{c}(v)\) for a vector \( v \in \mathbb{R}^d \) is defined as
\[
\exp_0^{c}(v) = \tanh\left(\sqrt{c} \|v\|\right) \frac{v}{\sqrt{c} \|v\|},
\]
with the convention \(\exp_0^{c}(0) = 0\). This projects euclidean vectors \(v\) onto the poincaré hyperbolic manifold in a way that preserves geometric structure. For each downstream NACSP task \( t \), we aggregate features from all subspaces \( k = 1, \ldots, K \) using attention weights \(\alpha_t(k)\) such that
\[
\sum_{k=1}^K \alpha_t(k) = 1, \quad \alpha_t(k) \geq 0.
\]
The attention mechanism allows the model to dynamically assign importance to each subspace \( z^{(k)} \) based on their relevance to the task \( t \), enabling selective focus on the most informative features. The aggregated poincaré representation \( h_t \in \mathbb{H}^d \) is computed using Möbius scalar multiplication \(\otimes_{c_k}\) and möbius addition \(\oplus_{c_k}\):
\[
h_t = \bigoplus_{k=1}^K \alpha_t(k) \otimes_{c_k} z^{(k)}.
\]
Möbius addition \(\oplus_c\) for \( x, y \in \mathbb{H}^d \) is defined as
\[
x \oplus_c y = \frac{(1 + 2c \langle x, y \rangle + c \|y\|^2) x + (1 - c \|x\|^2) y}{1 + 2c \langle x, y \rangle + c^2 \|x\|^2 \|y\|^2},
\]
where \(\langle x, y \rangle\) is the euclidean inner product and \(\|\cdot\|\) is the euclidean norm and möbius scalar multiplication \(\otimes_c\) for a scalar \( r \in \mathbb{R} \) and \( x \in \mathbb{H}^d \) is given by
\[
r \otimes_c x = \tanh\left(r \tanh^{-1}(\sqrt{c} \|x\|)\right) \frac{x}{\sqrt{c} \|x\|},
\]
with the convention \(r \otimes_c 0 = 0\) and where \(\tanh^{-1}\) denotes the inverse hyperbolic tangent. To reduce entanglement among the different poincaré latent factors \( z^{(k)} \), we propose a hyperbolic total correlation (HTC) loss defined as
\[
L_{\text{HTC}}^{\text{hyp}} = \sum_{i \neq j} D_{\mathrm{KL}}\bigl(p(z^{(i)}, z^{(j)}) \parallel p(z^{(i)}) p(z^{(j)})\bigr),
\]
which penalizes mutual information between pairs of hyperbolic latent representations, thereby promoting their statistical independence. For each downstream NACSP task \( t \), the aggregated poincaré representation \( h_t \in \mathbb{H}^d \) is then mapped back to the euclidean tangent space at the origin using the task-specific poincaré logarithmic map: 

\[h_t^{\text{tangent}} = \log_0^{c_t}(h_t) \] 
where
\[
\log_0^{c}(y) = \frac{2}{\sqrt{c}} \tanh^{-1}(\sqrt{c} \|y\|) \frac{y}{\|y\|},
\]
with the convention \(\log_0^{c}(0) = 0\). This mapping is performed individually for each task \( t \), enabling task-specific geometric decoding of the shared representation. The resulting vector \( h_t^{\text{tangent}} \in \mathbb{R}^d \) for each task is then passed through a task-specific fully connected network consisting of two hidden layers and a linear output layer to perform the corresponding NACSP task. \textbf{\texttt{HYDRA}} trainable parameters falls between 8-12M parameters depending on the input representation size of the PTMs.

\section{Experiments and Results}

\subsection{Dataset}
We employ two datasets in our experiments: ST-codecfake and CodecFake. For both the datasets, we label the fake audio corresponding to different NACs with the NAC parameters quantizers (Q), sampling rate (SR), and bits per second (BPS) respectively for NACSP.  \par

\noindent \textbf{ST-Codecfake} \cite{xie2025neural}: It is the first-dataset specifically built for source attribution of CFs. This dataset is constructed to evaluate NAC source attribution in both closed-set and open-set conditions. ST-Codecfake comprises bilingual audio samples synthesized using eleven SOTA NACs with Mimi, SpeechTokenizer, FunCodec, EnCodec, FACodec, SNAC as closed-set NACs and WavTokenizer, AcademicCodec, DAC,  AudioDec, SoundStream as open-set NACs. The real data is sourced from two corpora—VCTK and AISHELL-3. For training and validation, the dataset provides 70000 and 7000 audio samples respectively, each labeled across seven classes (one real and six codec-generated). The evaluation set comprises 158736 samples, including not only the closed-set NAC classes but also five open-set NACs. We use only the fake samples and follow the distribution given by them for the training and evaluation of the models. \par

\noindent \textbf{CodecFake} \cite{wu24p_interspeech}: It is one of the primary dataset proposed for building detection models for CFs. The dataset is built upon the VCTK corpus, leveraging its speaker diversity and recording consistency to simulate realistic scenarios of NAC-based speech synthesis. It makes use of EnCodec, SpeechTokenizer, AcademiaCodec, AudioDec, Descript Audio Codec and FunCodec. The authors has also explicitly mentioned the Q, SR, and BPS which are the three main NAC parameters NACSP estimates. For each NAC variant, the train, validation, and test subsets mirror the statistical distribution of the original VCTK source, resulting in 42752 training utterances, 735 validation utterances, and 755 test utterances per NAC subset. We use only the fake samples and the official training, validation, testing split for training and evaluation of the models.

\noindent \textbf{Training Details}: We train all the models for 50 epochs with batch size as 32 and learning of 1e-3. We use MSE (mean squared error) as loss function for all the three NACSP tasks (Q, BPS, SR) with Adam as the optimizer. We also make use of dropout and early stopping for prevention of overfitting.

\begin{table}[!h]
\setlength{\tabcolsep}{4pt}
\centering
\scriptsize
\begin{tabular}{l|cc|cc|cc}
\toprule
& \multicolumn{2}{c}{\textbf{SR}} & \multicolumn{2}{c}{\textbf{BPS}} & \multicolumn{2}{c}{\textbf{Q}} \\
\cmidrule(lr){2-3} \cmidrule(lr){4-5} \cmidrule(lr){6-7}
\textbf{PTM} & \textbf{RMSE $\downarrow$} & \textbf{MAE $\downarrow$} & \textbf{RMSE $\downarrow$} & \textbf{MAE $\downarrow$} & \textbf{RMSE $\downarrow$} & \textbf{MAE $\downarrow$} \\
\midrule
\multicolumn{7}{c}{\textbf{Baseline (Euclidean)}} \\
\midrule
WL & \cellcolor{blueaccent!15}6.89 & \cellcolor{blueaccent!15}5.43 & \cellcolor{blueaccent!15}2.75 & \cellcolor{blueaccent!15}1.89 & \cellcolor{blueaccent!15}6.36 & \cellcolor{blueaccent!15}5.51 \\
UN & \cellcolor{blueaccent!15}6.54 & \cellcolor{blueaccent!15}5.05 & \cellcolor{blueaccent!15}2.79 & \cellcolor{blueaccent!15}1.38 & \cellcolor{blueaccent!15}3.03 & \cellcolor{blueaccent!15}2.41 \\
W2 & \cellcolor{blueaccent!15}8.83 & \cellcolor{blueaccent!15}7.24 & \cellcolor{blueaccent!15}2.97 & \cellcolor{blueaccent!15}2.56 & \cellcolor{blueaccent!15}3.27 & \cellcolor{blueaccent!15}2.91 \\
XS & \cellcolor{blueaccent!15}7.87 & \cellcolor{blueaccent!15}6.42 & \cellcolor{blueaccent!15}3.38 & \cellcolor{blueaccent!15}2.44 & \cellcolor{blueaccent!15}6.81 & \cellcolor{blueaccent!15}5.27 \\
WS & \cellcolor{blueaccent!15}6.89 & \cellcolor{blueaccent!15}5.32 & \cellcolor{blueaccent!15}3.96 & \cellcolor{blueaccent!15}2.71 & \cellcolor{blueaccent!15}6.96 & \cellcolor{blueaccent!15}5.43 \\
MM & \cellcolor{blueaccent!15}8.65 & \cellcolor{blueaccent!15}7.61 & \cellcolor{blueaccent!15}3.31 & \cellcolor{blueaccent!15}2.59 & \cellcolor{blueaccent!15}8.30 & \cellcolor{blueaccent!15}7.36 \\
XT & \cellcolor{blueaccent!15}7.29 & \cellcolor{blueaccent!15}6.25 & \cellcolor{blueaccent!15}4.27 & \cellcolor{blueaccent!15}3.52 & \cellcolor{blueaccent!15}4.93 & \cellcolor{blueaccent!15}3.24 \\
EP & \cellcolor{blueaccent!15}6.94 & \cellcolor{blueaccent!15}5.19 & \cellcolor{blueaccent!15}3.44 & \cellcolor{blueaccent!15}2.62 & \cellcolor{blueaccent!15}4.29 & \cellcolor{blueaccent!15}3.34 \\
\midrule
\multicolumn{7}{c}{\textbf{Baseline (Hyperbolic)}} \\
\midrule
WL & \cellcolor{blueaccent!30}4.39 & \cellcolor{blueaccent!30}3.11 & \cellcolor{blueaccent!30}2.30 & \cellcolor{blueaccent!30}1.96 & \cellcolor{blueaccent!30}4.18 & \cellcolor{blueaccent!30}3.03 \\
UN & \cellcolor{blueaccent!30}4.26 & \cellcolor{blueaccent!30}3.02 & \cellcolor{blueaccent!30}2.25 & \cellcolor{blueaccent!30}1.81 & \cellcolor{blueaccent!30}3.42 & \cellcolor{blueaccent!30}2.58 \\
W2 & \cellcolor{blueaccent!30}4.76 & \cellcolor{blueaccent!30}3.48 & \cellcolor{blueaccent!30}2.36 & \cellcolor{blueaccent!30}2.09 & \cellcolor{blueaccent!30}3.51 & \cellcolor{blueaccent!30}2.78 \\
XS & \cellcolor{blueaccent!30}4.49 & \cellcolor{blueaccent!30}3.21 & \cellcolor{blueaccent!30}2.58 & \cellcolor{blueaccent!30}2.31 & \cellcolor{blueaccent!30}4.20 & \cellcolor{blueaccent!30}3.05 \\
WS & \cellcolor{blueaccent!30}4.35 & \cellcolor{blueaccent!30}3.07 & \cellcolor{blueaccent!30}2.81 & \cellcolor{blueaccent!30}2.48 & \cellcolor{blueaccent!30}4.25 & \cellcolor{blueaccent!30}3.14 \\
MM & \cellcolor{blueaccent!30}4.71 & \cellcolor{blueaccent!30}3.52 & \cellcolor{blueaccent!30}2.54 & \cellcolor{blueaccent!30}2.27 & \cellcolor{blueaccent!30}4.56 & \cellcolor{blueaccent!30}3.39 \\
XT & \cellcolor{blueaccent!30}4.42 & \cellcolor{blueaccent!30}3.19 & \cellcolor{blueaccent!30}3.04 & \cellcolor{blueaccent!30}2.63 & \cellcolor{blueaccent!30}3.81 & \cellcolor{blueaccent!30}2.96 \\
EP & \cellcolor{blueaccent!30}4.33 & \cellcolor{blueaccent!30}3.12 & \cellcolor{blueaccent!30}2.48 & \cellcolor{blueaccent!30}2.18 & \cellcolor{blueaccent!30}3.51 & \cellcolor{blueaccent!30}2.65 \\
\midrule
\multicolumn{7}{c}{\textbf{\texttt{HYDRA}}} \\
\midrule
WL & \cellcolor{blueaccent!50}3.09 & \cellcolor{blueaccent!50}1.87 & \cellcolor{blueaccent!50}1.09 & \cellcolor{blueaccent!50}1.00 & \cellcolor{blueaccent!50}2.88 & \cellcolor{blueaccent!50}2.78 \\
UN & \cellcolor{blueaccent!50}3.00 & \cellcolor{blueaccent!50}1.80 & \cellcolor{blueaccent!50}1.02 & \cellcolor{blueaccent!50}0.90 & \cellcolor{blueaccent!50}2.00 & \cellcolor{blueaccent!50}1.90 \\
W2 & \cellcolor{blueaccent!50}3.60 & \cellcolor{blueaccent!50}2.23 & \cellcolor{blueaccent!50}1.34 & \cellcolor{blueaccent!50}1.12 & \cellcolor{blueaccent!50}2.06 & \cellcolor{blueaccent!50}2.04 \\
XS & \cellcolor{blueaccent!50}3.35 & \cellcolor{blueaccent!50}2.07 & \cellcolor{blueaccent!50}1.33 & \cellcolor{blueaccent!50}1.30 & \cellcolor{blueaccent!50}3.00 & \cellcolor{blueaccent!50}2.71 \\
WS & \cellcolor{blueaccent!50}3.09 & \cellcolor{blueaccent!50}1.85 & \cellcolor{blueaccent!50}1.64 & \cellcolor{blueaccent!50}1.40 & \cellcolor{blueaccent!50}3.04 & \cellcolor{blueaccent!50}2.75 \\
MM & \cellcolor{blueaccent!50}3.56 & \cellcolor{blueaccent!50}2.30 & \cellcolor{blueaccent!50}1.36 & \cellcolor{blueaccent!50}1.29 & \cellcolor{blueaccent!50}3.40 & \cellcolor{blueaccent!50}3.30 \\
XT & \cellcolor{blueaccent!50}3.21 & \cellcolor{blueaccent!50}2.00 & \cellcolor{blueaccent!50}1.78 & \cellcolor{blueaccent!50}1.69 & \cellcolor{blueaccent!50}2.51 & \cellcolor{blueaccent!50}2.30 \\
EP & \cellcolor{blueaccent!50}3.13 & \cellcolor{blueaccent!50}1.94 & \cellcolor{blueaccent!50}1.23 & \cellcolor{blueaccent!50}1.20 & \cellcolor{blueaccent!50}2.21 & \cellcolor{blueaccent!50}2.21 \\
\bottomrule
\end{tabular}
\caption{Performance metrics for ST-Codecfake under closed-set settings where the NACs are seen during training: Abbreviations: WL = WavLM, UN = Unispeech-SAT, W2 = Wav2vec2, XS = XLS-R, WS = Whisper, MM = MMS, XT = x-vector, EP = ECAPA; The abbreviations used in this Table are kept same for Table \ref{codec} and \ref{st-codecout}}
\label{st-codec}
\end{table}

\begin{table}[hbt!]
\setlength{\tabcolsep}{4pt}
\centering
\scriptsize
\begin{tabular}{l|cc|cc|cc}
\toprule
& \multicolumn{2}{c}{\textbf{SR}} & \multicolumn{2}{c}{\textbf{BPS}} & \multicolumn{2}{c}{\textbf{Q}} \\
\cmidrule(lr){2-3} \cmidrule(lr){4-5} \cmidrule(lr){6-7}
\textbf{PTM} & \textbf{RMSE \(\downarrow\)} & \textbf{MAE \(\downarrow\)} & \textbf{RMSE \(\downarrow\)} & \textbf{MAE \(\downarrow\)} & \textbf{RMSE \(\downarrow\)} & \textbf{MAE \(\downarrow\)} \\
\midrule
\multicolumn{7}{c}{\textbf{Baseline (Euclidean)}} \\
\midrule
WL & \cellcolor{blueaccent!15}19.81 & \cellcolor{blueaccent!15}18.66 & \cellcolor{blueaccent!15}9.81  & \cellcolor{blueaccent!15}8.43  & \cellcolor{blueaccent!15}21.15 & \cellcolor{blueaccent!15}19.62 \\
UN & \cellcolor{blueaccent!15}20.58 & \cellcolor{blueaccent!15}19.48 & \cellcolor{blueaccent!15}8.96  & \cellcolor{blueaccent!15}7.57  & \cellcolor{blueaccent!15}19.39 & \cellcolor{blueaccent!15}18.70 \\
W2 & \cellcolor{blueaccent!15}19.39 & \cellcolor{blueaccent!15}18.25 & \cellcolor{blueaccent!15}8.76  & \cellcolor{blueaccent!15}7.37  & \cellcolor{blueaccent!15}21.25 & \cellcolor{blueaccent!15}19.81 \\
XS & \cellcolor{blueaccent!15}18.83 & \cellcolor{blueaccent!15}17.57 & \cellcolor{blueaccent!15}9.20  & \cellcolor{blueaccent!15}8.40  & \cellcolor{blueaccent!15}21.47 & \cellcolor{blueaccent!15}20.38 \\
WS & \cellcolor{blueaccent!15}19.97 & \cellcolor{blueaccent!15}18.86 & \cellcolor{blueaccent!15}9.11  & \cellcolor{blueaccent!15}8.58  & \cellcolor{blueaccent!15}20.62 & \cellcolor{blueaccent!15}19.21 \\
MM & \cellcolor{blueaccent!15}19.70 & \cellcolor{blueaccent!15}18.46 & \cellcolor{blueaccent!15}9.40  & \cellcolor{blueaccent!15}8.76  & \cellcolor{blueaccent!15}21.92 & \cellcolor{blueaccent!15}19.99 \\
XT & \cellcolor{blueaccent!15}19.57 & \cellcolor{blueaccent!15}18.32 & \cellcolor{blueaccent!15}8.85  & \cellcolor{blueaccent!15}7.34  & \cellcolor{blueaccent!15}20.58 & \cellcolor{blueaccent!15}19.67 \\
EP & \cellcolor{blueaccent!15}19.79 & \cellcolor{blueaccent!15}18.36 & \cellcolor{blueaccent!15}9.41  & \cellcolor{blueaccent!15}8.76  & \cellcolor{blueaccent!15}20.86 & \cellcolor{blueaccent!15}19.14 \\
\midrule
\multicolumn{7}{c}{\textbf{Baseline (Hyperbolic)}} \\
\midrule
WL & \cellcolor{blueaccent!30}13.04 & \cellcolor{blueaccent!30}10.89 & \cellcolor{blueaccent!30}13.42 & \cellcolor{blueaccent!30}11.05 & \cellcolor{blueaccent!30}13.26 & \cellcolor{blueaccent!30}10.87 \\
UN & \cellcolor{blueaccent!30}13.67 & \cellcolor{blueaccent!30}11.52 & \cellcolor{blueaccent!30}12.93 & \cellcolor{blueaccent!30}10.21 & \cellcolor{blueaccent!30}12.78 & \cellcolor{blueaccent!30}10.18 \\
W2 & \cellcolor{blueaccent!30}12.93 & \cellcolor{blueaccent!30}10.63 & \cellcolor{blueaccent!30}12.75 & \cellcolor{blueaccent!30}10.18 & \cellcolor{blueaccent!30}13.35 & \cellcolor{blueaccent!30}10.97 \\
XS & \cellcolor{blueaccent!30}12.66 & \cellcolor{blueaccent!30}10.44 & \cellcolor{blueaccent!30}13.09 & \cellcolor{blueaccent!30}11.18 & \cellcolor{blueaccent!30}13.49 & \cellcolor{blueaccent!30}11.46 \\
WS & \cellcolor{blueaccent!30}13.20 & \cellcolor{blueaccent!30}10.98 & \cellcolor{blueaccent!30}13.02 & \cellcolor{blueaccent!30}11.25 & \cellcolor{blueaccent!30}13.23 & \cellcolor{blueaccent!30}10.79 \\
MM & \cellcolor{blueaccent!30}13.03 & \cellcolor{blueaccent!30}10.89 & \cellcolor{blueaccent!30}13.17 & \cellcolor{blueaccent!30}11.31 & \cellcolor{blueaccent!30}13.61 & \cellcolor{blueaccent!30}11.30 \\
XT & \cellcolor{blueaccent!30}13.00 & \cellcolor{blueaccent!30}10.86 & \cellcolor{blueaccent!30}12.85 & \cellcolor{blueaccent!30}10.22 & \cellcolor{blueaccent!30}13.29 & \cellcolor{blueaccent!30}10.85 \\
EP & \cellcolor{blueaccent!30}13.14 & \cellcolor{blueaccent!30}10.93 & \cellcolor{blueaccent!30}13.21 & \cellcolor{blueaccent!30}11.32 & \cellcolor{blueaccent!30}13.41 & \cellcolor{blueaccent!30}10.99 \\
\midrule
\multicolumn{7}{c}{\textbf{\texttt{HYDRA}}} \\
\midrule
WL & \cellcolor{blueaccent!50}11.56 & \cellcolor{blueaccent!50}9.57  & \cellcolor{blueaccent!50}11.98 & \cellcolor{blueaccent!50}9.77  & \cellcolor{blueaccent!50}11.70 & \cellcolor{blueaccent!50}9.55 \\
UN & \cellcolor{blueaccent!50}12.19 & \cellcolor{blueaccent!50}10.32 & \cellcolor{blueaccent!50}11.21 & \cellcolor{blueaccent!50}9.16  & \cellcolor{blueaccent!50}11.02 & \cellcolor{blueaccent!50}9.13 \\
W2 & \cellcolor{blueaccent!50}11.31 & \cellcolor{blueaccent!50}9.36  & \cellcolor{blueaccent!50}11.04 & \cellcolor{blueaccent!50}9.05  & \cellcolor{blueaccent!50}11.72 & \cellcolor{blueaccent!50}9.64 \\
XS & \cellcolor{blueaccent!50}11.07 & \cellcolor{blueaccent!50}9.14  & \cellcolor{blueaccent!50}11.41 & \cellcolor{blueaccent!50}9.75  & \cellcolor{blueaccent!50}11.81 & \cellcolor{blueaccent!50}9.97 \\
WS & \cellcolor{blueaccent!50}11.63 & \cellcolor{blueaccent!50}9.68  & \cellcolor{blueaccent!50}11.37 & \cellcolor{blueaccent!50}9.89  & \cellcolor{blueaccent!50}11.52 & \cellcolor{blueaccent!50}9.34 \\
MM & \cellcolor{blueaccent!50}11.42 & \cellcolor{blueaccent!50}9.47  & \cellcolor{blueaccent!50}11.59 & \cellcolor{blueaccent!50}9.79  & \cellcolor{blueaccent!50}11.94 & \cellcolor{blueaccent!50}9.83 \\
XT & \cellcolor{blueaccent!50}11.38 & \cellcolor{blueaccent!50}9.45  & \cellcolor{blueaccent!50}11.09 & \cellcolor{blueaccent!50}9.02  & \cellcolor{blueaccent!50}11.63 & \cellcolor{blueaccent!50}9.29 \\
EP & \cellcolor{blueaccent!50}11.53 & \cellcolor{blueaccent!50}9.52  & \cellcolor{blueaccent!50}11.63 & \cellcolor{blueaccent!50}9.78  & \cellcolor{blueaccent!50}11.75 & \cellcolor{blueaccent!50}9.61 \\
\bottomrule
\end{tabular}
\caption{Performance metrics for CodecFake under closed-set settings where the NACs are seen during training}
\label{codec}
\end{table}

\subsection{Experimental Results}
As mentioned in Section \ref{intro}, NACSP combines both in-distribution and open-set scenarios into a unified task. However, for evaluation purposes, we assess our models separately on closed-set and open-set settings to better analyze the performance of NACSP. ST-Codecfake has both in-distribution and open-set evaluation, however, CodecFake has only in-distribution settings. We employ RMSE and MAE as our evaluation metrics, as these are among the most commonly used metrics in speech processing tasks.  Firstly, we present the comprehensive evaluation of the performance of various PTMs using the baseline modeling setup (Section \ref{baseline}). The results are presented in Table \ref{st-codec}, \ref{codec}, \ref{st-codecout} Baseline (Euclidean). A key observation from this evaluation is the absence of any consistently superior PTM across all NACSP tasks for both in-distribution and open-set settings. While certain PTMs demonstrate marginally better performance on specific tasks—for instance, some PTMs perform slightly better in estimating SR, while others show improved performance for BPS or Q—these variations are neither statistically significant nor consistent across the datasets (CodecFake and ST-CodecFake). This suggests that no single PTM inherently possesses a clear advantage over others in the context of NACSP tasks. Instead, the PTMs appear to encode comparable levels of task-relevant acoustic information, with minor differences. \par

\begin{figure}[hbt!]
    \centering
    \includegraphics[width=0.7\linewidth]{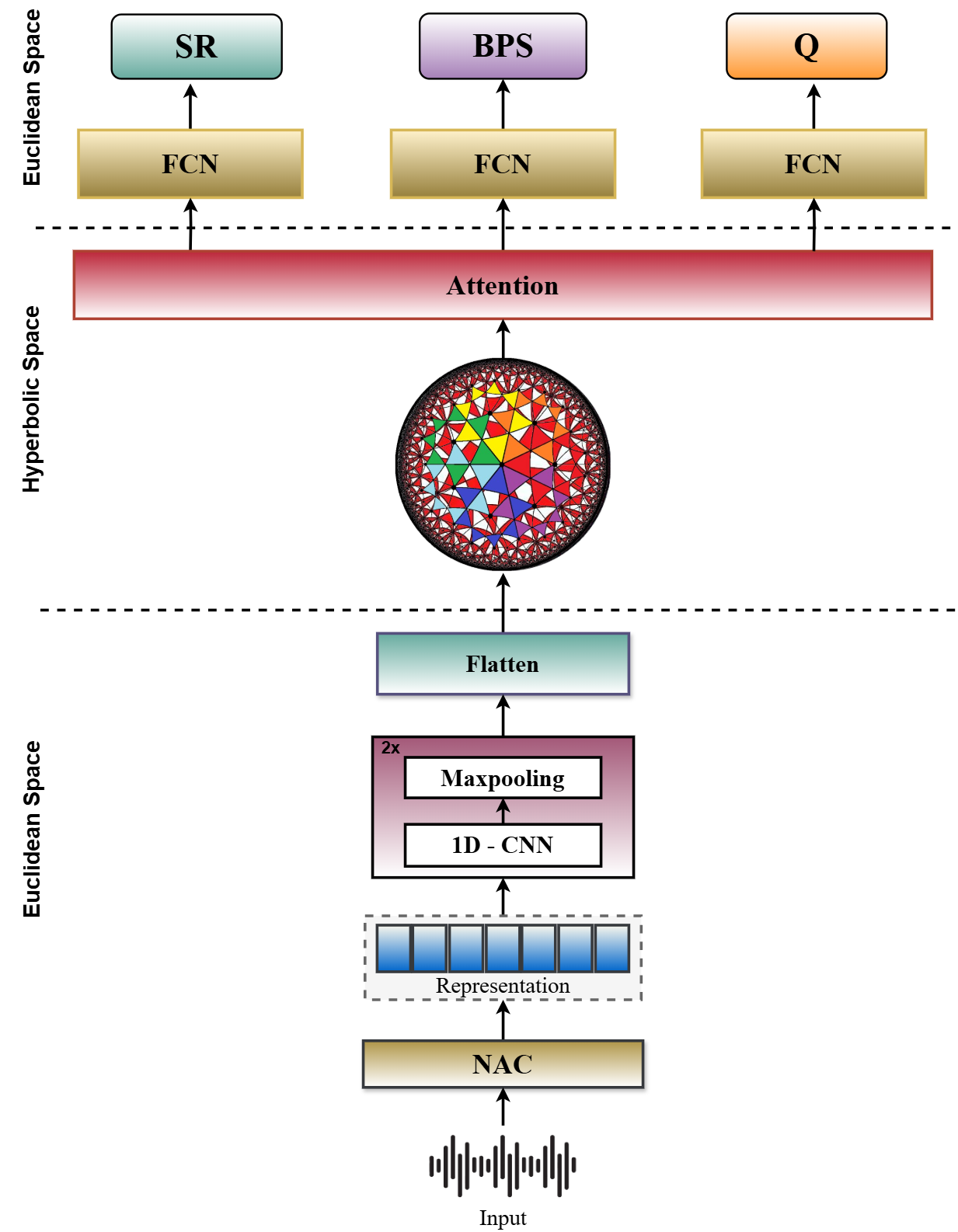}
    \caption{Baseline (Hyperbolic); FCN stands for fully connected network}
    \label{baselinehyperbolic}
\end{figure}

Then, we present the closed-set evaluation results using the proposed framework, \textbf{\texttt{HYDRA}}. These results, reported in Table \ref{st-codec} and Table \ref{codec}, clearly indicate that integrating PTMs within the \textbf{\texttt{HYDRA}} architecture yields substantially improved performance across all NACSP tasks—SR, BPS, and Q—compared to the baseline modeling setup. This improvement can be primarily attributed to task-wise hyperbolic disentanglement which enables the model to better isolate and model task-specific cues from the entangled PTM representations. Interestingly, the core finding from the baseline evaluation continues to hold even in this scenario. Despite the architectural improvements in terms of \textbf{\texttt{HYDRA}}, we still observe no substantial or consistent performance differentiation among the PTMs across the NACSP tasks. While \textbf{\texttt{HYDRA}} improves absolute performance across all tasks compared to the baseline—the relative performance rankings of the PTMs remain largely unchanged. That is, no single PTM emerges as clearly superior or consistently underperforms across the SR, BPS, and Q estimation tasks on both the CodecFake and ST-CodecFake datasets. This reinforces the earlier insight that the choice of PTM has limited impact in this context, and that most PTMs encode comparably useful acoustic features relevant to NACSP. \par

\begin{table}[hbt!]
\setlength{\tabcolsep}{4pt}
\centering
\scriptsize
\begin{tabular}{l|cc|cc|cc}
\toprule
& \multicolumn{2}{c}{\textbf{SR}} & \multicolumn{2}{c}{\textbf{BPS}} & \multicolumn{2}{c}{\textbf{Q}} \\
\cmidrule(lr){2-3} \cmidrule(lr){4-5} \cmidrule(lr){6-7}
\textbf{PTM} & \textbf{RMSE \(\downarrow\)} & \textbf{MAE \(\downarrow\)} & \textbf{RMSE \(\downarrow\)} & \textbf{MAE \(\downarrow\)} & \textbf{RMSE \(\downarrow\)} & \textbf{MAE \(\downarrow\)} \\
\midrule
\multicolumn{7}{c}{\textbf{Baseline (Euclidean)}} \\
\midrule
WL & \cellcolor{blueaccent!15}12.95 & \cellcolor{blueaccent!15}10.54 & \cellcolor{blueaccent!15}4.35 & \cellcolor{blueaccent!15}3.56 & \cellcolor{blueaccent!15}5.85 & \cellcolor{blueaccent!15}4.94 \\
UN & \cellcolor{blueaccent!15}12.84 & \cellcolor{blueaccent!15}10.39 & \cellcolor{blueaccent!15}4.41 & \cellcolor{blueaccent!15}3.72 & \cellcolor{blueaccent!15}5.97 & \cellcolor{blueaccent!15}5.02 \\
W2 & \cellcolor{blueaccent!15}13.66 & \cellcolor{blueaccent!15}12.15 & \cellcolor{blueaccent!15}4.79 & \cellcolor{blueaccent!15}3.99 & \cellcolor{blueaccent!15}6.13 & \cellcolor{blueaccent!15}5.21 \\
XS & \cellcolor{blueaccent!15}13.27 & \cellcolor{blueaccent!15}11.84 & \cellcolor{blueaccent!15}4.63 & \cellcolor{blueaccent!15}3.59 & \cellcolor{blueaccent!15}5.91 & \cellcolor{blueaccent!15}4.99 \\
WS & \cellcolor{blueaccent!15}12.42 & \cellcolor{blueaccent!15}11.10 & \cellcolor{blueaccent!15}4.37 & \cellcolor{blueaccent!15}3.65 & \cellcolor{blueaccent!15}5.88 & \cellcolor{blueaccent!15}4.97 \\
MM & \cellcolor{blueaccent!15}14.49 & \cellcolor{blueaccent!15}13.06 & \cellcolor{blueaccent!15}5.13 & \cellcolor{blueaccent!15}4.76 & \cellcolor{blueaccent!15}6.18 & \cellcolor{blueaccent!15}5.45 \\
XT & \cellcolor{blueaccent!15}12.69 & \cellcolor{blueaccent!15}13.29 & \cellcolor{blueaccent!15}4.54 & \cellcolor{blueaccent!15}3.55 & \cellcolor{blueaccent!15}5.91 & \cellcolor{blueaccent!15}4.62 \\
EP & \cellcolor{blueaccent!15}12.51 & \cellcolor{blueaccent!15}11.17 & \cellcolor{blueaccent!15}4.11 & \cellcolor{blueaccent!15}3.21 & \cellcolor{blueaccent!15}5.53 & \cellcolor{blueaccent!15}4.64 \\
\midrule
\multicolumn{7}{c}{\textbf{Baseline (Hyperbolic)}} \\
\midrule
WL & \cellcolor{blueaccent!30}7.12 & \cellcolor{blueaccent!30}5.91 & \cellcolor{blueaccent!30}4.08 & \cellcolor{blueaccent!30}3.34 & \cellcolor{blueaccent!30}4.52 & \cellcolor{blueaccent!30}3.73 \\
UN & \cellcolor{blueaccent!30}7.05 & \cellcolor{blueaccent!30}5.83 & \cellcolor{blueaccent!30}4.15 & \cellcolor{blueaccent!30}3.41 & \cellcolor{blueaccent!30}4.67 & \cellcolor{blueaccent!30}3.85 \\
W2 & \cellcolor{blueaccent!30}7.52 & \cellcolor{blueaccent!30}6.21 & \cellcolor{blueaccent!30}4.29 & \cellcolor{blueaccent!30}3.56 & \cellcolor{blueaccent!30}4.88 & \cellcolor{blueaccent!30}4.07 \\
XS & \cellcolor{blueaccent!30}7.30 & \cellcolor{blueaccent!30}6.08 & \cellcolor{blueaccent!30}4.22 & \cellcolor{blueaccent!30}3.50 & \cellcolor{blueaccent!30}4.69 & \cellcolor{blueaccent!30}3.91 \\
WS & \cellcolor{blueaccent!30}6.91 & \cellcolor{blueaccent!30}5.76 & \cellcolor{blueaccent!30}4.10 & \cellcolor{blueaccent!30}3.35 & \cellcolor{blueaccent!30}4.61 & \cellcolor{blueaccent!30}3.82 \\
MM & \cellcolor{blueaccent!30}7.94 & \cellcolor{blueaccent!30}6.61 & \cellcolor{blueaccent!30}4.35 & \cellcolor{blueaccent!30}3.63 & \cellcolor{blueaccent!30}5.02 & \cellcolor{blueaccent!30}4.18 \\
XT & \cellcolor{blueaccent!30}7.10 & \cellcolor{blueaccent!30}5.91 & \cellcolor{blueaccent!30}4.26 & \cellcolor{blueaccent!30}3.55 & \cellcolor{blueaccent!30}4.81 & \cellcolor{blueaccent!30}4.01 \\
EP & \cellcolor{blueaccent!30}6.98 & \cellcolor{blueaccent!30}5.79 & \cellcolor{blueaccent!30}4.17 & \cellcolor{blueaccent!30}3.46 & \cellcolor{blueaccent!30}4.58 & \cellcolor{blueaccent!30}3.79 \\
\midrule
\multicolumn{7}{c}{\textbf{\texttt{HYDRA}}} \\
\midrule
WL & \cellcolor{blueaccent!50}6.26 & \cellcolor{blueaccent!50}4.05 & \cellcolor{blueaccent!50}2.07 & \cellcolor{blueaccent!50}1.05 & \cellcolor{blueaccent!50}2.20 & \cellcolor{blueaccent!50}1.15 \\
UN & \cellcolor{blueaccent!50}6.20 & \cellcolor{blueaccent!50}4.00 & \cellcolor{blueaccent!50}2.09 & \cellcolor{blueaccent!50}1.07 & \cellcolor{blueaccent!50}2.27 & \cellcolor{blueaccent!50}1.19 \\
W2 & \cellcolor{blueaccent!50}6.60 & \cellcolor{blueaccent!50}4.61 & \cellcolor{blueaccent!50}2.20 & \cellcolor{blueaccent!50}1.10 & \cellcolor{blueaccent!50}2.37 & \cellcolor{blueaccent!50}1.28 \\
XS & \cellcolor{blueaccent!50}6.41 & \cellcolor{blueaccent!50}4.50 & \cellcolor{blueaccent!50}2.15 & \cellcolor{blueaccent!50}1.05 & \cellcolor{blueaccent!50}2.23 & \cellcolor{blueaccent!50}1.18 \\
WS & \cellcolor{blueaccent!50}6.00 & \cellcolor{blueaccent!50}4.24 & \cellcolor{blueaccent!50}2.08 & \cellcolor{blueaccent!50}1.06 & \cellcolor{blueaccent!50}2.22 & \cellcolor{blueaccent!50}1.17 \\
MM & \cellcolor{blueaccent!50}6.92 & \cellcolor{blueaccent!50}5.00 & \cellcolor{blueaccent!50}2.25 & \cellcolor{blueaccent!50}1.20 & \cellcolor{blueaccent!50}2.40 & \cellcolor{blueaccent!50}1.36 \\
XT & \cellcolor{blueaccent!50}6.15 & \cellcolor{blueaccent!50}4.89 & \cellcolor{blueaccent!50}2.18 & \cellcolor{blueaccent!50}1.19 & \cellcolor{blueaccent!50}2.31 & \cellcolor{blueaccent!50}1.24 \\
EP & \cellcolor{blueaccent!50}6.09 & \cellcolor{blueaccent!50}4.12 & \cellcolor{blueaccent!50}2.12 & \cellcolor{blueaccent!50}1.08 & \cellcolor{blueaccent!50}2.16 & \cellcolor{blueaccent!50}1.12 \\
\bottomrule
\end{tabular}
\caption{Performance metrics for ST-Codecfake under open-set settings where the NACs are not seen during training}
\label{st-codecout}
\end{table}

A key takeaway from our results for future researchers is that while the selection of a particular PTM is not critical, substantial performance gains can be achieved by integrating PTMs within our proposed framework, \textbf{\texttt{HYDRA}}. Importantly, this observation holds true not only in the closed-set evaluation but also in the open-set setting, as reported in Table~\ref{st-codecout}. In both scenarios, \textbf{\texttt{HYDRA}} consistently achieves top performance across all NACSP tasks—SR, BPS, and Q—highlighting its robustness and effectiveness in modeling NAC-specific traits. Notably, the relative invariance in PTM performance persists even in the open-set scenario, reinforcing the conclusion that the specific choice of PTM has minimal impact on overall NACSP outcomes. To further investigate the importance of architectural components in \textbf{\texttt{HYDRA}}, we conducted an ablation study (Shown in Figure \ref{baselinehyperbolic}) where the shared encoder representations were projected into a single hyperbolic space instead of distinct task-specific ones, followed by task-wise prediction heads. We also removed the attention modules and the HTC loss, while keeping the rest of the architectural details identical to \textbf{\texttt{HYDRA}}. Although this variant outperformed the Baseline (Euclidean) through both closed-set and open-set setting, it is no match to \textbf{\texttt{HYDRA}}. This clearly demonstrates the benefit of using task-specific hyperbolic subspaces for effective NACSP.

\begin{table}[htbp]
\centering
\scriptsize
\begin{tabular}{l|cc|cc|cc}
\toprule
\textbf{Model} & \multicolumn{2}{c|}{\textbf{SR}} & \multicolumn{2}{c|}{\textbf{BPS}} & \multicolumn{2}{c}{\textbf{Q}} \\
\cmidrule(lr){2-3} \cmidrule(lr){4-5} \cmidrule(lr){6-7}
              & \textbf{RMSE} & \textbf{MAE} & \textbf{RMSE} & \textbf{MAE} & \textbf{RMSE} & \textbf{MAE} \\
\midrule
\multicolumn{7}{c}{\textbf{ST-Codecfake (CS)}} \\
\midrule
\textbf{H}     & \textbf{3.21} & \textbf{2.00} & \textbf{1.78} & \textbf{1.69} & \textbf{2.51} & \textbf{2.30} \\
M     & 7.95 & 6.60 & 6.65 & 5.32 & 7.37 & 6.10 \\
A     & 8.11 & 7.66 & 8.79 & 6.45 & 8.55 & 7.21 \\
\midrule
\multicolumn{7}{c}{\textbf{ST-Codecfake (OS)}} \\
\midrule
\textbf{H}     & \textbf{6.15} & \textbf{4.89} & \textbf{2.18} & \textbf{1.19} & \textbf{2.31} & \textbf{1.24} \\
M     & 10.42 & 9.78 & 7.05 & 6.10 & 7.49 & 6.03 \\
A     & 9.67 & 8.03 & 6.21 & 5.24 & 6.64 & 5.17 \\
\midrule
\multicolumn{7}{c}{\textbf{CodecFake}} \\
\midrule
\textbf{H}     & \textbf{11.38} & \textbf{9.45} & \textbf{11.09} & \textbf{9.02} & \textbf{11.63} & \textbf{9.29} \\
M     & 15.65 & 13.21 & 16.48 & 14.83 & 17.02 & 16.11 \\
A     & 17.01 & 15.58 & 16.91 & 15.20 & 18.43 & 17.56 \\
\bottomrule
\end{tabular}
\caption{Comparison to SOTA models; CS: Closed-set, OS: Open-set, H: \texttt{\textbf{HYDRA(x-vector)}}, M: MiO (Whisper + x-vector), A: AASIST (Wav2vec2)}
\label{sota_comparison_table}
\end{table}

\noindent \textbf{Comparison with SOTA}:  As NACSP is novel task, so for better understanding and comparing our proposed framework \textbf{\texttt{HYDRA}}, we reimplemented two well known SOTA modeling architectures for ADD \cite{chetia-phukan-etal-2024-heterogeneity} and source attribution \cite{klein24_interspeech}. \citealt{chetia-phukan-etal-2024-heterogeneity} used combination of Whisper and x-vector PTM representations with bilinear-pooling, while \citealt{klein24_interspeech} used AASIST with Wav2vec2 representations. We observe that \textbf{\texttt{HYDRA}} with x-vector gives the best performance in comparison to previous SOTA methods. Thus, setting SOTA in NACSP in both closed-set and open-set settings. The results are shown in Table \ref{sota_comparison_table}. Here, we have used x-vector with \textbf{\texttt{HYDRA}} as we have reported that there is no particular difference in performance with usage of different PTMs. 

\section{Conclusion}

In conclusion, we introduced, NACSP, a paradigm shift in addressing the limitations of existing open-set attribution approaches for CFs. Unlike conventional classification-based methods that merely flag unknown sources as generic ``unknowns'', NACSP provides a structured and fine-grained alternative by estimation of the underlying NAC parameters—such as Q, BPS, and SR. By estimating the parameters, we can trace back to the exact NAC that generated a particular audio. We also proposed \textbf{\texttt{HYDRA}}, a novel hyperbolic space-based framework that enables effective disentanglement of latent PTM representations through task-specific attention across multiple curvature-aware subspaces. Our comprehensive evaluation across multiple benchmarks demonstrates that \textbf{\texttt{HYDRA}} significantly outperforms traditional euclidean baselines, setting a new SOTA in both closed-set and open-set settings. These findings underscore the importance of moving beyond discrete attribution towards structured regression for more fine-grained and generalizable forensic insights into NAC generation.

\bibliography{main}

\end{document}